\documentclass[12pt]{article}
\usepackage{savesym}
\usepackage{amsfonts}
\usepackage{stmaryrd}

\usepackage{amsmath}
\savesymbol{iint}
\usepackage{txfonts}
\restoresymbol{TXF}{iint}

\usepackage{bbold}
\usepackage{latexsym}
\usepackage{enumerate}

\newdimen\proofrulebreadth \proofrulebreadth=.05em
\newdimen\proofdotseparation \proofdotseparation=1.25ex
\newdimen\proofrulebaseline \proofrulebaseline=2ex
\newcount\proofdotnumber \proofdotnumber=3
\let\then\relax
\def\hfi{\hskip0pt plus.0001fil}
\mathchardef\squigto="3A3B
\newif\ifinsideprooftree\insideprooftreefalse
\newif\ifonleftofproofrule\onleftofproofrulefalse
\newif\ifproofdots\proofdotsfalse
\newif\ifdoubleproof\doubleprooffalse
\let\wereinproofbit\relax
\newdimen\shortenproofleft
\newdimen\shortenproofright
\newdimen\proofbelowshift
\newbox\proofabove
\newbox\proofbelow
\newbox\proofrulename
\def\shiftproofbelow{\let\next\relax\afterassignment\setshiftproofbelow\dimen0 }
\def\shiftproofbelowneg{\def\next{\multiply\dimen0 by-1 }%
\afterassignment\setshiftproofbelow\dimen0 }
\def\setshiftproofbelow{\next\proofbelowshift=\dimen0 }
\def\setproofrulebreadth{\proofrulebreadth}

\def\prooftree{
\ifnum  \lastpenalty=1
\then   \unpenalty
\else   \onleftofproofrulefalse
\fi
\ifonleftofproofrule
\else   \ifinsideprooftree
        \then   \hskip.5em plus1fil
        \fi
\fi
\bgroup
\setbox\proofbelow=\hbox{}\setbox\proofrulename=\hbox{}%
\let\justifies\proofover\let\leadsto\proofoverdots\let\Justifies\proofoverdbl
\let\using\proofusing\let\[\prooftree
\ifinsideprooftree\let\]\endprooftree\fi
\proofdotsfalse\doubleprooffalse
\let\thickness\setproofrulebreadth
\let\shiftright\shiftproofbelow \let\shift\shiftproofbelow
\let\shiftleft\shiftproofbelowneg
\let\ifwasinsideprooftree\ifinsideprooftree
\insideprooftreetrue
\setbox\proofabove=\hbox\bgroup$\displaystyle 
\let\wereinproofbit\prooftree
\shortenproofleft=0pt \shortenproofright=0pt \proofbelowshift=0pt
\onleftofproofruletrue\penalty1
}

\def\eproofbit{
\ifx    \wereinproofbit\prooftree
\then   \ifcase \lastpenalty
        \then   \shortenproofright=0pt  
        \or     \unpenalty\hfil         
        \or     \unpenalty\unskip       
        \else   \shortenproofright=0pt  
        \fi
\fi
\global\dimen0=\shortenproofleft
\global\dimen1=\shortenproofright
\global\dimen2=\proofrulebreadth
\global\dimen3=\proofbelowshift
\global\dimen4=\proofdotseparation
\global\count255=\proofdotnumber
$\egroup  
\shortenproofleft=\dimen0
\shortenproofright=\dimen1
\proofrulebreadth=\dimen2
\proofbelowshift=\dimen3
\proofdotseparation=\dimen4
\proofdotnumber=\count255
}

\def\proofover{
\eproofbit 
\setbox\proofbelow=\hbox\bgroup 
\let\wereinproofbit\proofover
$\displaystyle
}%
\def\proofoverdbl{
\eproofbit 
\doubleprooftrue
\setbox\proofbelow=\hbox\bgroup 
\let\wereinproofbit\proofoverdbl
$\displaystyle
}%
\def\proofoverdots{
\eproofbit 
\proofdotstrue
\setbox\proofbelow=\hbox\bgroup 
\let\wereinproofbit\proofoverdots
$\displaystyle
}%
\def\proofusing{
\eproofbit 
\setbox\proofrulename=\hbox\bgroup 
\let\wereinproofbit\proofusing
\kern0.3em$
}

\def\endprooftree{
\eproofbit 
  \dimen5 =0pt
\dimen0=\wd\proofabove \advance\dimen0-\shortenproofleft
\advance\dimen0-\shortenproofright
\dimen1=.5\dimen0 \advance\dimen1-.5\wd\proofbelow
\dimen4=\dimen1
\advance\dimen1\proofbelowshift \advance\dimen4-\proofbelowshift
\ifdim  \dimen1<0pt
\then   \advance\shortenproofleft\dimen1
        \advance\dimen0-\dimen1
        \dimen1=0pt
        \ifdim  \shortenproofleft<0pt
        \then   \setbox\proofabove=\hbox{%
                        \kern-\shortenproofleft\unhbox\proofabove}%
                \shortenproofleft=0pt
        \fi
\fi
\ifdim  \dimen4<0pt
\then   \advance\shortenproofright\dimen4
        \advance\dimen0-\dimen4
        \dimen4=0pt
\fi
\ifdim  \shortenproofright<\wd\proofrulename
\then   \shortenproofright=\wd\proofrulename
\fi
\dimen2=\shortenproofleft \advance\dimen2 by\dimen1
\dimen3=\shortenproofright\advance\dimen3 by\dimen4
\ifproofdots
\then
        \dimen6=\shortenproofleft \advance\dimen6 .5\dimen0
        \setbox1=\vbox to\proofdotseparation{\vss\hbox{$\cdot$}\vss}%
        \setbox0=\hbox{%
                \advance\dimen6-.5\wd1
                \kern\dimen6
                $\vcenter to\proofdotnumber\proofdotseparation
                        {\leaders\box1\vfill}$%
                \unhbox\proofrulename}%
\else   \dimen6=\fontdimen22\the\textfont2 
        \dimen7=\dimen6
        \advance\dimen6by.5\proofrulebreadth
        \advance\dimen7by-.5\proofrulebreadth
        \setbox0=\hbox{%
                \kern\shortenproofleft
                \ifdoubleproof
                \then   \hbox to\dimen0{%
                        $\mathsurround0pt\mathord=\mkern-6mu%
                        \cleaders\hbox{$\mkern-2mu=\mkern-2mu$}\hfill
                        \mkern-6mu\mathord=$}%
                \else   \vrule height\dimen6 depth-\dimen7 width\dimen0
                \fi
                \unhbox\proofrulename}%
        \ht0=\dimen6 \dp0=-\dimen7
\fi
\let\doll\relax
\ifwasinsideprooftree
\then   \let\VBOX\vbox
\else   \ifmmode\else$\let\doll=$\fi
        \let\VBOX\vcenter
\fi
\VBOX   {\baselineskip\proofrulebaseline \lineskip.2ex
        \expandafter\lineskiplimit\ifproofdots0ex\else-0.6ex\fi
        \hbox   spread\dimen5   {\hfi\unhbox\proofabove\hfi}%
        \hbox{\box0}%
        \hbox   {\kern\dimen2 \box\proofbelow}}\doll%
\global\dimen2=\dimen2
\global\dimen3=\dimen3
\egroup 
\ifonleftofproofrule
\then   \shortenproofleft=\dimen2
\fi
\shortenproofright=\dimen3
\onleftofproofrulefalse
\ifinsideprooftree
\then   \hskip.5em plus 1fil \penalty2
\fi
}

\newcommand{\id}{{\rm id}}

\renewcommand{\to}{\longrightarrow}

\newcommand{\LLL}{{\cal L}}

\newcommand{\TTT}{{\cal T}}

\renewcommand{\Bbb}{\mathbb}

\newcommand{\CCc}{{\Bbb C}}

\mathcode`\<="4268 
\mathcode`\>="5269 
\mathchardef\gt="313E 
\mathchardef\lt="313C 
\newsavebox{\barr}
\savebox{\barr}{\hspace*{-9.5pt}\raisebox{1.25pt}{$\scriptscriptstyle%
|$}\hspace*{4.5pt}} 
\newsavebox{\barrleft}
\savebox{\barrleft}{\hspace*{-8.5pt}\raisebox{1.25pt}{$\scriptscriptstyle%
|$}\hspace*{10pt}}

 \def\pushright#1{{
    \parfillskip=0pt            
    \widowpenalty=10000         
    \displaywidowpenalty=10000  
    \finalhyphendemerits=0      
    \leavevmode                 
    \unskip                     
    \nobreak                    
    \hfil                       
    \penalty50                  
    \hskip.2em                  
    \null                       
    \hfill                      
    {#1}                        
    \par}}                      

 \def\qed{\pushright{$\square$}\penalty-700 \smallskip}

\newenvironment{prf}[1]{\begin{trivlist} \item[{\bf ~Proof}#1.]}%
{\qed\end{trivlist}}

\newcommand{\beq}{\begin{equation}}
\newcommand{\eeq}{\end{equation}}
\newcommand{\ba}[1]{\begin{array}{#1}}
\newcommand{\ea}{\end{array}}
\newcommand{\bea}{\begin{eqnarray}}
\newcommand{\eea}{\end{eqnarray}}
\newcommand{\bear}{\begin{eqnarray*}}
\newcommand{\eear}{\end{eqnarray*}}
\newcommand{\bpr}{\begin{prf}{}}
\newcommand{\epr}{\end{prf}}
\newcommand{\bprf}[1]{\begin{prf}{#1}}
\newcommand{\eprf}{\end{prf}}

\usepackage{xy}
\xyoption{all}
\CompileMatrices

\usepackage{code}

\newcommand{\lbcat}{\mbox{$[\negmedspace ($}}
\newcommand{\rbcat}{\mbox{$)\negmedspace ]$}}

\newcommand{\ana}[1]{\mbox{$\lbcat #1 \rbcat$}}
\newcommand{\cata}[1]{\llparenthesis #1 \rrparenthesis}
\newcommand{\sana}[1]{[\negmedspace ( #1 )\negmedspace ]}

\renewcommand{\to}{\longrightarrow}
\newcommand{\edge}[2]{\xymatrix @-1.2pc{#1\ar@{)}[r]& #2}}
\newcommand{\eedge}[2]{\def\objectstyle{\scriptstyle}
\xymatrix @-1.5pc{#1\ar@{)}[r]& #2}}
\newcommand{\eeedge}[3]{\xymatrix @-1pc{#1\ar@{)}[r]^-{#3}& #2}}
  
\newcommand{\reduces}{\blacktriangleright\blacktriangleright}
\newcommand{\comp}{\cdot}
\newcommand{\PN}{{\sf Para}}

\newcommand{\und}{\raisebox{-.6ex}{-}}

\def\tms{$\times$}
\newcommand{\bld}{{\tt build}}

\newtheorem{theorem}{Theorem}[section]
\newtheorem{proposition}[theorem]{Proposition}
\newtheorem{lemma}[theorem]{Lemma}

\newtheorem{definition}[theorem]{Definition}

\begin{document}

\title{Logic of fusion\\[1ex]
\emph{\large --- Dedicated to Andre Scedrov  ---}}

\author{Dusko Pavlovic\thanks{Supported by NSF and AFOSR.}\\
University of Hawaii, Honolulu HI\\
{\small dusko@hawaii.edu}
}

\date{}

\date{}

\maketitle

\section*{Personal introduction}
I first met Andre at one of the Peripatetic Seminars on Sheaves and Logic (PSSL) in the late 80s. PSSL was a legendary community of category theorists, meeting a couple of times a year at venerable universities in Europe and the UK. Andre was a well-established researcher, who had already subsumed forcing under the classifying topos construction; and I was a wide-eyed grad student. He was pointed out to me as coming from the same country where I had come from (which at the time still existed); but the main reason why I had already read maybe not all, but most of his papers, was that I was trying to learn and understand the powerful new methods of category theory that Andre had worked on. 

Nowadays, you probably wouldn't call either Andre or me a category theorist. The word ''category'' does not occur that often either in his or in my papers. Yet, if you follow the common thread that ties together Andre's work, it takes you through logic, the semantics of computation, decision procedures and algorithms, models of natural language, security protocols. It is a very long thread. Longer than just a thread of good taste, of clever constructions, and honest excitement. It is a thread of method. By trying to trace this thread through Andre's work, I confront the challenge of explaining my own. How did we get from toposes and constructivist universes to distance bounding protocols and mafia attacks?  Of course I don't know the answer.  One answer might be that we got wiser. Another answer is that we are still too young to tell. Or is that just a wise way to avoid answering?

Instead of attempting to answer, or to avoid answering, I recall an intermediary step. I reproduce for the occasion a construction from a long time ago, that arose when I read \cite{scedrov1990guide}, and then \cite{BainbridgeES:funp,carboni1987categorical,FreydPJ:semppl}. The construction was never published, although it indirectly led to \cite{PavlovicD:FOSSACS01}. It was developed for specific applications in a tool that I was trying to build \cite{PavlovicD:ARSPA06}, but the conceptual problem was reduced to the toy task of polymorphic zipping. By that time, Andre was already past the polymorphism research phase. When I caught up with him at the next corner, we were both thinking about security. There seems to be some sort of polymorphism behind it all.

\section{Technical introduction} 
\subsection{Idea}
The starting point of this work is the observation that the Curry-Howard isomorphism \cite{Curry-festschrift}, relating
\begin{eqnarray*}
\mbox{types} &\leftrightsquigarrow & \mbox{propositions}\\
\mbox{programs} &\leftrightsquigarrow & \mbox{proofs}\\
\mbox{composition} &\leftrightsquigarrow & \mbox{cut}
\end{eqnarray*}
can be extended by a correspondence of
\begin{eqnarray*}
\mbox{\bf program fusion} &\mathbf\leftrightsquigarrow & \mbox{\bf cut
elimination} 
\end{eqnarray*}
This simple idea suggests logical interpretations of some of the basic
methods of generic and transformational programming. In the present
paper, we provide a logical analysis of the general form of {\em build
fusion}, also known as {\em deforestation}, over the inductive and the
coinductive datatypes, regular or nested. The analysis is based on a
logical reinterpretation of parametricity \cite{strachey1967fundamental} in terms of {\em
paranatural\/} transformations, modifying the functorial interpretation of polymorphism in \cite{BainbridgeES:funp}.

\subsection{Fusion and cut}\label{Sec:cut}
The Curry-Howard isomorphism is one of the conceptual building blocks
of type theory, built deep into the foundation of computer science and
functional programming \cite[Ch.~3]{GirardJY:prot}. The fact that it
is an {\em isomorphism} means that the type and the term constructors
on one side obey the same laws as the logical connectives, and the
logical derivation ruleson the other side. For instance, the products
and the sums of types correspond, respectively, to the conjunction and
the disjunction, because the respective introduction rules
\begin{eqnarray*}
\prooftree A\vdash B_0 \quad A\vdash B_1 
\justifies A \vdash B_0 \wedge B_1 
	\using{\wedge I}
\endprooftree &\hspace{5em}& 
\prooftree A_0\vdash B\quad A_1\vdash B 
\justifies A_0\vee A_1 \vdash B
	\using{\vee I}
\endprooftree
\end{eqnarray*}
extended by the labels for proofs, yield the type formation rules
\begin{eqnarray*}
\prooftree
		f_0: A\rightarrow B_0  \quad
		f_1: A\rightarrow B_1
	\justifies
		<f_0, f_1>: A \rightarrow B_0 \times B_1
\endprooftree
&\hspace{5em}& 
\prooftree
		g_0: A_0\rightarrow B  \quad
		g_1: A_1\rightarrow B
	\justifies
		[g_0, g_1]: A_0 + A_1 \rightarrow B\endprooftree
\end{eqnarray*}
In a sense, the pairing constructors $<-,->$ and $[-,-]$ record on the
terms the applications of the rules $\wedge I$ and $\vee I$, as the proof
constructors.

Extending this line of thought a step further, one notices that the
term reductions also mirror the proof transformations. E.g., the
transformation
\begin{eqnarray*}
\prooftree
	\prooftree 
		A_0 \vdash B \quad
		A_1 \vdash B
	\justifies
		A_0 \vee A_1 \vdash B
	\endprooftree
	B\vdash C
\justifies
	A_0 \vee A_1 \vdash C
\endprooftree
&\hspace{3em} \reduces\hspace{2.5em}&
\prooftree
	\prooftree 
		A_0 \vdash B \qquad
		B \vdash C
	\justifies
		A_0 \vdash C
	\endprooftree
	\prooftree 
		A_1 \vdash B \qquad
		B \vdash C
	\justifies
		A_1 \vdash C
	\endprooftree
\justifies
	A_0 \vee A_1 \vdash C
\endprooftree
\end{eqnarray*}
corresponds to the rewrite
\begin{eqnarray}\label{rewrite}
h\comp [f_0,f_1] &\ \ \  \reduces\ \  & [h\comp f_0\ ,\ h\comp f_1]
\end{eqnarray}
where $f_0$ and $f_1$ are the labels of the proofs $A_0\vdash B$
and $A_1 \vdash B$, whereas $h$ is the label of $B\vdash C$. The
point of such transformations is that the applications of
the cut rule
\begin{equation}
\prooftree
		A \vdash B \qquad
		B \vdash C
	\justifies
		A \vdash C
\endprooftree
\end{equation}
get pushed up the proof tree, as to be eliminated, by iterating
such moves. On the side of terms and programs, the cut, of course,
corresponds to the composition
\begin{equation}\label{old}
\prooftree
		f: A \rightarrow B \qquad
		h: B \rightarrow C
	\justifies
		h\comp f: A \rightarrow C
\endprooftree
\end{equation}
Just like the presence of a cut in a proof means that an intermediary
proposition has been created, and then cut out, the presence of the
composition in a program means that the thread of computation leads through an intermediary type, used to pass data between the components, and then discarded. Computational aspects of normalization are discussed in \cite[Ch.~4]{GirardJY:prot}.

While the programs decomposed into simple parts are easier to write
and understand, passing the data and control between the components
incurs a computational overhead. For instance, running the composite
$\tt ssum\comp zipW$ of
\begin{code}
	zipW               : [Nat]{\tms}[Nat] -> [Nat{\tms}Nat]
	zipW (x::xs,y::ys) = (x,y) :: zip xs ys
	zipW  (xs,   ys)   = []\end{code}
and
\begin{code}
	ssum           : [Nat{\tms}Nat] -> Nat 
	ssum []        = 0 
	ssum (x,y)::zs = x + y + sum zs\end{code} 
is clearly less efficient than running the fusion
\begin{code}
	sumzip                : [Nat]{\tms}[Nat] -> Nat
	sumzip (x::xs,y::ys)  =  x + y + sumzip (xs,ys)
	sumzip  (xs,   ys)    =  0\end{code}
where the intermediary lists {\tt [Nat{\tms}Nat]} are eliminated. In
	practice, the 	data structures passed between the 
components tend to be very large, and the gain by eliminating them can be
significant. On the other hand, the efficient, monolythic code,
obtained by fusion, tends to be more complex, and thus harder to
understand and maintain.

To get both efficiency and compositionality, to allow the programmers to
write simple, modular code, and optimize it in the compilation, the program
fusions need to be sufficiently well understood to be automated. Our
first point is that the Curry-Howard isomorphism maps this task onto
the well-ploughed ground of logic.

\subsection{Build fusion}\label{Sec:fusion}
The general form of the build fusion that we shall study corresponds, in
the inductive case, to the ``cut
rule''
\begin{equation}\label{build}
\prooftree
\xymatrix @-.5pc{\vspace{2ex} & \\A \ar[r]^{f} & M_F}\quad\quad
\xymatrix @-.5pc{FM_F\ar[r]^-{F\cata{c}}\ar[d]_{\mu} & FC\ar[d]^{c}\\
M_F \ar[r]^{\cata{c}} & C}  
\justifies
\xymatrix @-.2pc{A \ar[rr]^{f'C (\ulcorner c\urcorner)} && C}
\endprooftree
\end{equation}
eliminating the inductive datatype $M_F$, which is the initial algebra
of the type constructor $F$.  In practice and in literature, $F$ is
usually a list- or a tree-like constructor, and the type $A$ is
often inductively defined itself; but we shall see that the above scheme is valid in its full generality. The {\tt sumzip}-example from
the preceding section can be obtained as an instance of this scheme,
taking $FX = 1+{\tt Nat}\times{\tt Nat}\times X$, and thus $M_F =
{\tt [Nat\times Nat]}$. The function {\tt ssum} is the catamorphism
(fold) of the map $[0,\ddag] : 1+{\tt Nat\times Nat\times Nat} \to
{\tt Nat}$ where $\ddag$ maps $<i,j,k>$ to $i+j+k$.

The dual scheme
\beq\label{buildd}
\prooftree
\def\objectstyle{\scriptstyle}\def\arrowstyle{\scriptscriptstyle}
\xymatrix{FA\ar[r]^-{F\sana{a}}
& FN_F\\   
A \ar[u]^-{a}  \ar[r]^{\sana{a}} &N_F \ar[u]_\nu}  
\quad\quad \xymatrix {\vspace{2ex} & \\ N_F
\ar[r]^-{g} & C}
\justifies\def\objectstyle{\scriptstyle}\def\arrowstyle{\scriptscriptstyle}
\xymatrix{A \ar[rr]^-{g'A (\ulcorner a\urcorner)} && C} 
\endprooftree
\eeq
allows eliminating the coinductive type $N_F$, the final $F$-coalgebra. 

Clearly, the essence of both of the above fusion schemes lies in the
terms $f'$ and $g'$. Where do they come from? The idea is to represent
the fixpoints $M_F$ and $N_F$ in their ``logical form''
\begin{eqnarray}
\label{RP}
M_F & \cong & \forall  X.\ (F X\Rightarrow  X)\Rightarrow X\\
\label{Has}
N_F & \cong & \exists X.\ X\times (X\Rightarrow FX) 
\end{eqnarray}
The parametric families
\begin{eqnarray}
f' X &: & (F X\Rightarrow  X) \to (A \Rightarrow X)\label{f'}\\
g' X & : & (X\Rightarrow FX)\  \to\ (X\Rightarrow C)\label{g'}
\end{eqnarray} 
are then obtained by extending $f:A\to M_F$ and $g:N_F\to C$ along
isomorphisms (\ref{RP}) and (\ref{Has}), and rearranging the
arguments. The equations
\begin{eqnarray}
\cata{c}\comp f &=& f' C(\ulcorner c\urcorner)\label{fst}\\
g\comp\ana{a} & = & g' A (\ulcorner a\urcorner)\label{snd}
\end{eqnarray} 
can be proved using logical relations, or their convenient derivative,
Wadler's ``theorems for free'' \cite{WadlerP:TFF}. This was indeed
done already in \cite{LaunchburyJ:shortcut}  for (\ref{fst}), and
(\ref{snd}) presents no problems either.

However, mapped along the Curry-Howard isomorphism, equations
(\ref{fst}--\ref{snd}) become statements about the equivalence of
proofs. The fact that all logical relations on all Henkin models must
relate the terms involved in these equations does not seem to offer a
clue for understanding their equivalence.

\subsection*{Overview of the paper}
In order to acquire some insight into the logical grounds of program
fusion, and equivalence, we propose {\em paranatural\/}
transformations, presented in Sec.~\ref{Paranaturals}, as a conceptually justified and technically useful instance of the dinatural semantics of polymorphism  \cite{BainbridgeES:funp}. The applicability of this concept is based on the characterization of the parametricity of families (\ref{f'}) and (\ref{g'}) in terms of an intrinsic
commutativity property. We note that this characterization is completely intrinsic, with  no recourse to models or external structures. The upshot is that the results actually apply much more widely than presented here, i.e. beyond the scope of build fusion. But that was the application that motivated the approach, and it suffices to show the case. The paranaturality condition is a variation on the theme of functorial and structural polymorphism \cite{BainbridgeES:funp,carboni1987categorical,FreydPJ:semppl,FreydPJ:strp,scedrov1990guide}. Unfortunately, neither of these semantical frameworks provides sufficient guidance for actual programming applications. The dinatural transformations of \cite{BainbridgeES:funp,FreydPJ:semppl} provide a conceptually clear view of polymorphism as an invariance property; but it has been recognized early on that the characterization is too weak, as it allows too many terms. On the other hand, the structor morphisms of \cite{FreydPJ:strp} precisely correspond to the accepted polymorphic terms; but the approach is not effective, as it does not stipulate which of the many possible choices of structors should be used to interpret a particular polytype. We propose paranatural transformations as a means for filling this gap. This proposal emerged from practical applications in programming. It is based on the insight, on the logical background of Propositions~\ref{prop:decomp} and \ref{prop:TTT},  that program fusion only ever requires capturing as polymorphic one of two kinds of families of computations: 
\begin{itemize}
\item those where the inputs \emph{from}\/ some final datatypes are consumed, and 
\item those where the outputs are produced {\em into\/} some initial datatypes.
\end{itemize}
Prop.~\ref{MN} in Sec.~\ref{Fixpoints} formalizes this idea. The proof of this proposition is given in the Appendix. The proofs of the other propositions are straightforward.  We note that the result eliminates the extensionality and the well-pointedness requirements of logical relations, which hamper their applications, even on the toy examples discussed here. On the other hand, refining the logical approach from Sections ~\ref{Sec:cut} and \ref{Sec:fusion} along the lines of \cite{PavlovicD:mapsII} seems to broaden the presented methods
beyond their current scope. Some evidence of this is discussed in the final section.

\section{Paranatural transformations}\label{Paranaturals}
As it has been well known at least since Freyd's work on recursive
types in algebraically compact categories \cite{FreydPJ:algcom},
separating the covariant and the contravariant occurrences of $X$ in a
polytype $\TTT(X)$ yields a polynomial functor
$T:\CCc^{op}\times\CCc\to \CCc$. On the other hand, by simple
structural induction, one easily proves that
\begin{proposition}\label{prop:decomp}
For every polynomial functor $T:\CCc^{op}\times \CCc\to \CCc$ over a
cartesian closed category $\CCc$, there are polynomial functors
$W:\CCc^{op}\times \CCc\to \CCc$ and $V:\CCc\to \CCc$, unique up to
isomorphism, such that
\begin{eqnarray*}
T & \cong &  W\Rightarrow V
\end{eqnarray*}
\end{proposition}
This motivates the following

\begin{definition}
Let $\CCc$ be a category and $W:\CCc^{op}\times \CCc\to C$ and
$V:\CCc\to \CCc$ functors on it.

A\/ {\em paranatural} transformation $\vartheta: W \to V$ is a family of
$\CCc$-arrows $\vartheta X : WXX\to VX$, such that for every arrow
$u:X\to Y$ in $\CCc$, the external pentagon in the following diagram
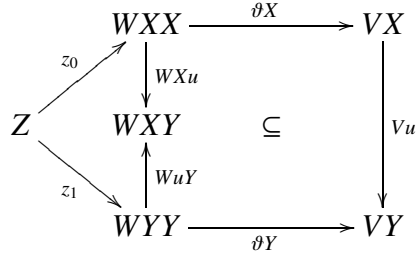
\[\xymatrix{
& WXX \ar[rr]^{\vartheta X} \ar[d]^{WXu} && VX \ar[dd]^{Vu}\\
Z\ar[ru]^{z_0} \ar[rd]_{z_1} & WXY & \subseteq\\
& WYY \ar[rr]_{\vartheta Y} \ar[u]_{WuY} && VY
}
\]
commutes whenever the triangle on the left commutes, for all $Z$, $z_0$
and $z_1$ in $\CCc$. This conditional commutativity is annotated by the $\subseteq$ inside the diagram.

The class of the paranatural transformations from $W$ to $V$ is
written $\PN(W,V)$.
\end{definition}

\noindent{\bf Remark.} When $\CCc$ supports calculus of relations, the quantification over $Z$, $z_0$ and $z_1$ and the entire triangle on the left can be omitted: the definition boils down to the requirement that the square commutes up to
$\subseteq$, in the relational sense.

\begin{proposition}\label{prop:TTT}
Let $\cal L$ be a polymorphic $\lambda$-calculus, and $\CCc_\LLL$ the
cartesian closed category generated by its closed types and terms. For
every type constructor $\TTT$, definable in $\cal L$, there is a
bijective correspondence
\begin{eqnarray*}
\CCc_\LLL\left(A,\ \forall X. \TTT(X)\right)&\cong & \PN(A\times W, V)
\end{eqnarray*}
natural in $A$.
\end{proposition}

\section{Characterizing fixpoints}\label{Fixpoints}
\begin{proposition}\label{MN}
Let $\CCc$ be a cartesian closed category, and $F$ a strong
endofunctor on it. Whenever the initial $F$-algebra $M_F$, resp. the
final $F$-coalgebra $N_F$ exist, then the following correspondences
hold
\begin{eqnarray}
\CCc(A, M_F) & \cong & \PN\left(A\times (FX\Rightarrow X),\
X\right)\label{bld1}\\   
\CCc(N_F, B) & \cong & \PN\left(X\times (X\Rightarrow FX),\
B\right)\label{bld2} 
\end{eqnarray}
naturally in $A$, resp. $B$.
\end{proposition}

The proof of this proposition is given in the Appendix.

In well-pointed categories and strongly extensional $\lambda$-calculi,
this proposition boils down to the following ``yoneda'' lemmas.

\noindent{\bf Notation.} Given $h:A\times B \to C$ and $b :1\to B$, we
write $h(b)$ for the result of partially evaluating $h$ on $b$
\[
\xymatrix{
A \ar[drr]_{h(b)} \ar[rr]^{<\id, b_!>} &&
A\times B \ar[d]^h \\  
&& C
}
\]
where $b_!$ denotes the composite $A \stackrel{!}{\rightarrow} 1
\stackrel{b}{\rightarrow} B$.

\begin{lemma}\label{M}
For paranatural transformations
\begin{eqnarray*}
\varphi_X & :&   A\times (FX\Rightarrow X) \ \to\  X\\
\psi_Y & :&   Y\times (Y\Rightarrow FY) \ \to \  B
\end{eqnarray*} 
hold the equations
\begin{eqnarray}\label{phi}
\varphi_X (\ulcorner x\urcorner) & = & \cata{x}\comp
\varphi_{M_F}(\mu)\\
\label{psi}
\psi_Y (\ulcorner y\urcorner) & = & \psi_{N_F}(\nu) \comp\ana{y} 
\end{eqnarray}
for all  $x:FX\to X$ and $y:Y\to FY$. 
\end{lemma}
While (\ref{phi}) follows from
\[\xymatrix{
& A\times FM_F \Rightarrow M_F \ar[rr]^-{\varphi M_F} \ar[d]^{A\times
FM_F\Rightarrow \cata{x}} && M_F \ar[dd]^{\cata{x}}\\
A\ar[ru]^{<\id, \ulcorner \mu\urcorner_!>} \ar[rd]_{<\id, \ulcorner
x\urcorner_! >} & A\times FM_F \Rightarrow X & \subseteq\\
& A\times FX\Rightarrow X \ar[rr]_-{\varphi X} \ar[u]_{A\times
F\cata{x}\Rightarrow X} && X
}
\]
(\ref{psi}) is obtained by chasing
\[\xymatrix{
& Y\times Y \Rightarrow FY \ar[rr]^-{\psi Y} \ar[d]^{\sana{y} \times
Y\Rightarrow F\sana{y}}&& B \ar[dd]^{\id} \\
Y\ar[ru]^-{<\id, \ulcorner y\urcorner_!>} \ar[rd]_-{<\sana{y}, \ulcorner
\nu \urcorner_! >} \ar[d]_{\sana{y}} & N_F\times Y \Rightarrow FN_F &
\subseteq \\
N_F \ar[r]_-{<\id, \ulcorner\nu\urcorner_!>} & N_F \times
N_F\Rightarrow FN_F \ar[rr]_-{\psi N_F} \ar[u]_{N_{F} \times 
\sana{y}\Rightarrow FN_F} && B
}
\]

In well-pointed categories, $\varphi_X : A\times (FX\Rightarrow X) \to
X$ is completely determined by its values $\varphi_X(\ulcorner
x\urcorner): A \to C$ on all $x:FX \to X$. Similarly, $\psi_Y :
Y\times (Y\Rightarrow FY) \to B$ is completely determined by its
values on $y: Y\to FY$.

However, in order to show that $\varphi_{M_F}(\mu)$ is generic for
$\varphi$ and $\psi_{N_F}(\nu)$ for $\psi$ without the
well-pointedness assumption, one needs to set up slightly different
constructions.

\section{Applications}
Using correspondence (\ref{bld1}), i.e. the maps realizing it, we can
now, first of all, provide the rational reconstruction of the simple
fusion from the introduction. The abstract form of the function {\tt
zipW} will be
\begin{code}
	zipW' : [Nat]{\tms}[Nat] -> ((1+Nat{\tms}Nat{\tms}X)->X)->X
	zipW' X (x::xs,y::ys) [m,c] = c(x, y, zipWith' X (xs,ys)
	[m,c]) zipW' X (xs, ys) [m,c] = m\end{code} 
While {\tt zipW} can be recovered as the instance \cd{zipW' [Nat{\tms}Nat] _ [[],(::)]}, 
i.e. {\tt zipW = build(zipW')}, the fusion is obtained as
\begin{code}
	sumzip  =  zipW' Nat _ [0,\ddag]\end{code}
But what is {\tt zipW}, if it is not a catamorphism? How come that it still
has a recursive definition?

It is in fact an {\em ana\/}morphism, and ${\tt ssum\comp zipW}$ can
be simplified by the coinductive build fusion as well. The scheme is
this time
\[
\prooftree
\def\objectstyle{\scriptstyle}\def\arrowstyle{\scriptscriptstyle}
\xymatrix{{\tt 1+ Nat\times Nat\times [Nat]\times [Nat]}\ar[r]
& {\tt 1+ Nat\times Nat \times [Nat\times Nat]}\\   
{\tt [Nat]\times [Nat]} \ar[u]^-{{\tt zW}}  \ar[r]^{{\tt zipW}} & {\tt
[Nat\times Nat]}\ar[u]}  
\quad\quad \xymatrix {\vspace{2ex} & \\ {\tt [Nat\times Nat]}
\ar[r]^-{{\tt ssum}} & {\tt Nat}}
\justifies\def\objectstyle{\scriptstyle}\def\arrowstyle{\scriptscriptstyle}
\xymatrix{{\tt [Nat]\times [Nat]} \ar[rrr]^-{{\tt ssum'\ [Nat]\times
[Nat]\ \und\  
zW}} &&& {\tt Nat}} 
\endprooftree
\]
where
\begin{code}
	zW (x::xs,y::ys) = (x,y,xs,ys)
	zW (xs,ys)	 = One \mbox{ (the element of 1)}\end{code}
induces ${\tt zipW} = \ana{{\tt zW}}$, whereas
\begin{code}
	ssum'       : X \tms (X -> 1+Nat{\tms}Nat{\tms}X ) -> Nat
	ssum' X x d = case d x of
			  One     -> 0
                          (n,m,y) -> n + m + ssum' X y d\end{code}
Calculating the conclusion this time yields
\begin{code}
	sumzip = ssum' [Nat]{\tms}[Nat] _ zW\end{code} Finally,
lifting proposition \ref{MN} to the category $\CCc^\CCc$ of
endofunctors, we can derive the build fusion rule for nested data
types \cite{BirdR:nest}. Consider, e.g., the type constructor {\tt
Nest}, that can be defined as a fixpoint of the functor
$\Psi:\CCc^\CCc \to \CCc^\CCc$, mapping 
$\Psi(F) = \lambda X. 1+ X\times F(X\times X)$.

The elements of the datatype {\tt Nest Nat} are the lists
where the $i$-th entry is an element of ${\tt Nat}^{2^i}$.
Abbreviating  {\tt Nest Nat} to {\tt \{Nat\}}, we can now define
\begin{code}
	zWN (x::xs,y::ys) = (x,y,fst xs,fst ys,
                                 snd xs,snd ys)
	zWN (xs,ys)	  = One\end{code}
where {\tt fst} and {\tt snd} are the obvious projections $\tt
	\{X\times X\} \to \{X\}$, and
and derive $\tt zipWN : \{Nat\}\times\{Nat\} \to \{Nat\times Nat\}$ as
	$\ana{{\tt zWN}}$ again. 
On the other hand, working out the paranaturality condition in
	$\CCc^\CCc$ allows 
	lifting 
\begin{code}
	ssumN           : \{Nat{\tms}Nat\} -> Nat 
	ssumN []        = 0 
	ssumN (x,y)::zs = x + y + ssumN (fst zs) 
                                + ssumN (snd zs)\end{code}
to
\begin{code}
	ssumN'         : F(Nat) \tms 
	                 F(X) -> 1+X{\tms}X{\tms}F(X{\tms}X) -> Nat 
	ssumN' F X f d = case d Nat f of
	       One     -> 0
               (n,m,g) -> m + n + ssumN' FF X g dd\end{code}
where {\tt FF} and {\tt dd} are the instances with {\tt X{\tms}X}
instead of {\tt X}. The fusion
\begin{code}
	sumzipN = ssumN' Nest{\tms}Nest Nat _ zWN\end{code}
is this time
\begin{code}
	sumzipN                : \{Nat\}{\tms}\{Nat\} -> Nat
	sumzipN (x::xs,y::ys)  =  x + y + sumzipN (fst xs,fst ys) +
				          sumzipN (snd xs,snd ys)
	sumzipN  (xs,   ys)    =  0\end{code}

\section{Afterword}	
The real application that motivated the presented work was a network application, based on event-channel architecture. A process involved a stream producer and a stream consumer, and the problem was to move filtering from the client side to the server side. Build fusion made this possible. The intermediary datatype, eliminated through build fusion, was thus infinitary: the streams. While the presented approach achieved its goal, and significantly improved the system, albeit in exchange for a lengthy derivation, the server at hand was actually a service aggregator, and thus also a client of other servers; and those servers were for their part also other servers' clients. So there was a cascade of streams to be eliminated by means of a cascade of build fusions. The upshot is that the theoretical approach presented here simplified the practical application; but the practical application demonstrated that the calculations needed to apply the theory were intractably complex. The task of automating the approach opened up, and remained open. On the bright side, the event-channels involved security protocols. As I was trying to learn more about that, I realized that structural methods seemed to apply in that area as well, and that it was under active explorations by Andre Scedrov, with many friends and collaborators \cite{Scedrov-Kanovich-Chadha,LincolnMMS99,ScedrovCGWW01}.

\bibliographystyle{abbrv}
\bibliography{poly,PavlovicD,logic}

\appendix 

\section*{Appendix: Proof of Prop.~\ref{MN}}

Towards isomorphism (\ref{bld1}), we define the maps 
\begin{eqnarray*}
(-)' &:& \CCc(A, M_F) \to \PN\left(A\times (FX\Rightarrow X),\
X\right)\\
\bld &:& \PN\left(A\times (FX\Rightarrow X),\ X\right)  \to \CCc(A, M_F)   
\end{eqnarray*}
and show that they are inverse to each other.

Given $f: A \to M_F$, the $X$-th component of $f'$ will be
\begin{eqnarray*}
f'_X\ : \ A\times (FX\Rightarrow X) & \stackrel{f\times k}{\to}&
M_F \times (M_F\Rightarrow X)\\ 
&\stackrel{\varepsilon}{\to} & X
\end{eqnarray*}
where $k : (FX\Rightarrow X) \to (M_F\Rightarrow X)$ maps the algebra
structures $x:FX\rightarrow X$ to the catamorphisms $\cata{x}:M_F
\rightarrow X$. Formally, $k$ is obtained by
transposing the catamorphism $\cata{\kappa}: M_F \to (FX\Rightarrow
X)\Rightarrow X$ for the $F$-algebra $\kappa$ on $(FX\Rightarrow
X)\Rightarrow X$, obtained by transposing the composite
\begin{eqnarray*}
\lefteqn{(FX\Rightarrow X)\times F\left((FX\Rightarrow X)\Rightarrow
X\right) \to}\\ 
&\stackrel{\rm (i)}\to & (FX\Rightarrow X)\times(FX\Rightarrow X)\times
F\left((FX\Rightarrow X)\Rightarrow X\right)\\
&\stackrel{\rm (ii)}\to & (FX\Rightarrow X)\times F\left((FX\Rightarrow
X)\times(FX\Rightarrow X)\Rightarrow X\right) \\
&\stackrel{\rm (iii)}\to & (FX\Rightarrow X)\times FX \\
&\stackrel{\rm (iv)}\to & X
\end{eqnarray*}
where arrow (i) is derived from the diagonal on $FX\Rightarrow X$,
(ii) from the strength, while (iii) and (iv) are just evaluations.

Towards the definition of $\bld$, for a paranatural $\varphi\ :\
A\times (FX\Rightarrow X) \to X$ take
\begin{eqnarray*}
\bld (\varphi)\ : \ A & \xymatrix @-.5pc{\ar[r]^{A\times \ulcorner\mu\urcorner_!} &}& A\times (FM_F \Rightarrow M_F)\\ 
& \xymatrix @-.5pc{\ar[r]^{\varphi M_F} &} & M_F
\end{eqnarray*}

Composing the above two definitions, one gets the commutative square
\[
\xymatrix{A \ar[dd]_{\bld (f')} \ar[rr]^-{A\times\ulcorner \mu
\urcorner_!} && A\times (FM_F\Rightarrow M_F) \ar[ddll]^{f' {M_F}}
\ar[dd]^{f\times k} \\
\\
M_F && M_F \times (M_F\Rightarrow M_F) \ar[ll]^\varepsilon
}
\] 
Since $k\comp\ulcorner \mu \urcorner = \ulcorner \id_M
\urcorner$, the path around the square reduces to $f$, and yields
$\bld (f') =  f$.

The converse $\bld (\varphi)' = \varphi$ is the point-free version
of lemma \ref{M}. It amounts to proving that the paranaturality of
$\varphi$ implies (indeed, it is equivalent) to the commutativity of
\[
\xymatrix{A \ar[dd]_{\widetilde{\varphi} X} \ar[rr]^-{A\times\ulcorner \mu
\urcorner_!} && A\times (FM_F\Rightarrow M_F) 
\ar[dd]^{\varphi M_F} \\
\\
(FX\Rightarrow X)\Rightarrow X && M_F \ar[ll]^{\cata{\kappa}}
}
\] 
where $\widetilde{\varphi}X$ is the transpose of $\varphi X$. Showing
this is an exercise in cartesian closed structure. On the other hand,
the path around the square is easily seen to be
$\bld (\varphi)'_X$.

To establish isomorphism (\ref{bld2}), we internalize \ref{psi} similarly
like we did \ref{phi} above. The natural correspondences
\begin{eqnarray*}
(-)' &:& \CCc(N_F,B) \to \PN\left(X\times (X\Rightarrow FX),\
B\right)\\
\bld &:& \PN\left(X\times (X\Rightarrow FX),\ B\right)  \to \CCc(N_F, B)   
\end{eqnarray*}
are defined
\begin{eqnarray*}
g'_X\ : \ X\times (X\Rightarrow FX) & \stackrel{X\times \ell}{\to}&
X \times (X\Rightarrow N_F)\\ 
&\stackrel{\varepsilon}{\to} & N_F\\
&\stackrel{g}{\to} & B
\end{eqnarray*}
and 
\begin{eqnarray*}
\bld (\psi)\ : \ N_F & \xymatrix @-.5pc{\ar[r]^{N_F\times
\ulcorner\nu\urcorner_!} &}& N_F \times (N_F \Rightarrow FN_F)\\ 
& \xymatrix @-.5pc{\ar[r]^{\psi N_F} &} & B
\end{eqnarray*}
for $g:N_F\to B$ and $\psi : X\times (X\Rightarrow FX) \to B$.
The arrow $\ell : (X\Rightarrow FX) \to (X\Rightarrow FX)$ maps the coalgebra
structures $x:X\rightarrow FX$ to the anamorphisms $\ana{x}:X
\rightarrow N_F$. \hfill $\Box$

\end{document}